*Alex Amadori*
*Eva Behrens*
*Gabriel Alfour*
*Andrea Miotti*

# The three main doctrines on the future of AI

# The three main doctrines on the future of AI

*Alex Amadori • Eva Behrens • Gabriel Alfour • Andrea Miotti*

Abstract:

This paper develops a taxonomy of expert perspectives on the risks and likely consequences of artificial intelligence, with particular focus on Artificial General Intelligence (AGI) and Artificial Superintelligence (ASI). Drawing from primary sources, we identify three predominant doctrines: (1) the dominance doctrine, which predicts that the first actor to create sufficiently advanced AI will attain overwhelming strategic superiority sufficient to cheaply neutralize its opponents' defenses; (2) the extinction doctrine, which anticipates that humanity will likely lose control of ASI, leading to the extinction of the human species or its permanent disempowerment; and (3) the replacement doctrine, which forecasts that AI will automate a large share of tasks currently performed by humans, but will not be so transformative as to fundamentally reshape or bring an end to human civilization. We examine the assumptions and arguments underlying each doctrine, including expectations around the pace of AI progress and the feasibility of maintaining advanced AI under human control. While the boundaries between doctrines are sometimes porous and many experts hedge across them, this taxonomy clarifies the core axes of disagreement over the anticipated scale and nature of the consequences of AI development.

# Contents

# 1. Introduction

AI systems are progressing rapidly, with frontier models being released every few months. The field has experienced major breakthroughs in recent years: the development of GPT-3 marked a stark inflection point by demonstrating the effectiveness of scaling large language models.[1] This was soon followed by another breakthrough, Reinforcement Learning from Human Feedback, which enabled the development of conversational AI systems like ChatGPT[2]. More recently, the field witnessed the emergence of reasoning models[3], which demonstrate superior performance over previous approaches in many tasks such as coding, mathematics and science.

Evidence suggests that AI systems are surpassing human experts across many fields. OpenAI's o3 model achieved competitive programming scores matching the 175th top contestant among 150,000 participants[4]. More recently, an undisclosed OpenAI model has achieved second place at the AtCoder Heuristics World Finals.[5] Research by METR indicates that AI systems' ability to autonomously complete complex tasks is advancing exponentially, with a doubling time of seven months, improving from 30-second tasks in March 2023 to 15-minute tasks by February 2025.[6]

In this backdrop, divergent views have emerged about both the trajectory of the technology itself and its potential impact on the world. Many experts anticipate that we will soon develop AI systems that can automate most tasks performed by humans; such systems are commonly referred to as Artificial General Intelligence (AGI).[7][8][9] Some expect this will quickly be followed by the development of AI surpassing the brightest human minds across all domains, or even surpassing the combined intelligence of humanity — what experts call Artificial Superintelligence (ASI). Notably, the number of efforts explicitly aiming for ASI has

---

[1] [Huge foundation models are turbo-charging AI progress](#), The Economist
[2] [Training a Helpful and Harmless Assistant with Reinforcement Learning from Human Feedback](#), Yuntao Bai et al.
[3] [Introducing OpenAI o1-preview](#), OpenAI
[4] [OpenAI o3 performance](#), OpenAI
[5] [OpenAI on X](#), OpenAI: Our model took 2nd place at the AtCoder Heuristics World Finals! Congrats to the champion for holding us off this time.
[6] [Measuring AI Ability to Complete Long Tasks](#), Thomas Kwa, Ben West, et al.
[7] [Three observations](#), Sam Altman: Systems that start to point to AGI* are coming into view
[8] [The Urgency of Interpretability](#), Dario Amodei: We could have AI systems equivalent to a 'country of geniuses in a datacenter' as soon as 2026 or 2027.
[9] [FAQ on Catastrophic AI Risks](#), Yoshua Bengio: My current estimate places a 95% confidence interval for the time horizon of super-human intelligence at 5 to 20 years.

increased markedly; this includes both research companies mainly or solely dedicated to developing ASI[10][11] as well as established tech companies like Meta and Alibaba.[12][13]

Yet significant disagreement persists about whether these milestones are achievable in the near term. Likewise, there is intense debate about the magnitude of their potential impact, with predictions ranging from the establishment of a golden age of human flourishing to catastrophic outcomes such as the extinction of the human species. In this paper, we survey expert statements on these topics, with particular emphasis on primary sources, and study the implication of these divergent views on considerations around national and global security. Through this lens, we observe that these perspectives fit naturally into three categories.

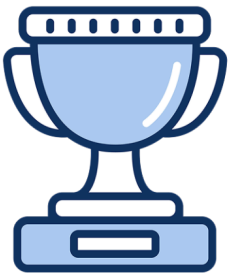

**Dominance doctrine.** The first actor to develop sufficiently advanced AI will gain overwhelming power over all others.

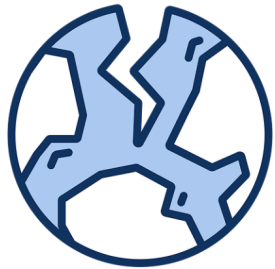

**Extinction doctrine.** Superintelligent AI will escape human control, leading to the extinction of humanity or its permanent disempowerment.

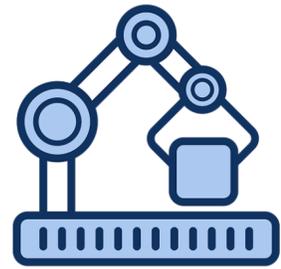

**Replacement doctrine.** AI promises greatly accelerated scientific and economic progress, but poses risks of extreme power concentration and mass manipulation.

**The dominance doctrine**, which predicts that the first actor to achieve ASI will gain a "decisive strategic advantage" over all other actors; that is, a position of strategic superiority sufficient to allow the actor to achieve overwhelming military and economic dominance over the rest of the world.

This doctrine expects that ASI will be developed in the near future, with some forecasting dates as early as 2026 to 2029. These systems will be able to accelerate two key domains.

---

[10] Safe Superintelligence Inc., Ilya Sutskever, Daniel Gross, Daniel Levy: We have started the world's first straight-shot SSI lab, with one goal and one product: a safe superintelligence.
[11] Reflections, Sam Altman: We are now confident we know how to build AGI as we have traditionally understood it … We are beginning to turn our aim beyond that, to superintelligence in the true sense of the word.
[12] Meta Is Creating a New A.I. Lab to Pursue 'Superintelligence', New York Times
[13] Alibaba CEO Wu says AGI is now company's primary objective, Bloomberg

The first is AI R&D itself; once a sufficient advantage is gained in this area, the gap would become self-reinforcing, making it impossible for competitors to catch up. The second is the research, development, and large-scale operation of military technologies. The combination of these two capabilities means that the first actor to develop sufficiently advanced AI may gain a strategic advantage over all other actors so vast that it creates the possibility of executing strikes that would neutralize all of their defenses in a relatively cheap and risk-free manner. After achieving this position, such an actor may be able to maintain an unassailable world order.

**The extinction doctrine**, which predicts that humanity will lose control over ASI, likely leading to its extinction or permanent disempowerment. A wide range of pathways to this outcome have been theorized, ranging from ones where a single, monolithic superintelligent AI system suddenly performs a hostile takeover after initially acting cooperative, to ones where control is gradually ceded to AI systems as they systematically replace humans across all economic, political and social functions.

While the extinction doctrine largely agrees with the dominance doctrine regarding the expected pace and extent of AI development, its core thesis is that we are not on track to develop techniques to maintain control of ASI by the time we develop such systems. Given how powerful superintelligent AI systems are projected to be, it would be impossible to maintain or regain control of them once they are pursuing goals incompatible with human values and interests, eventually leading to catastrophic outcomes such as the extinction or permanent subjugation of all of humanity.

**The replacement doctrine**, a broad umbrella of views predicting that AI development will result in AI replacing humans in carrying out some or most of the tasks they currently perform, allowing them to be executed at much higher speed and scale and more cheaply. However, they generally agree that AI development will not usher in radically new capabilities that may overturn existing economic and geopolitical paradigms. In more concrete terms, AI will not enable the development of military technology that allows one actor to prevail over all others; nor will it bring about catastrophic outcomes like human extinction. Being the most heterogeneous of the three, the views of proponents of the replacement doctrine span from highly optimistic to pessimistic about AI's impact on society.

There is disagreement within this school of thought over to what extent humans will be replaced in their roles as producers of economic value, with some expecting only partial automation of human labor and others envisioning near-complete replacement of humans in the economy. What characterizes the replacement doctrine is that even those anticipating the highest degrees of automation expect that AI will not be so transformative that we can no longer meaningfully speak of an "economy" existing, and that humans will keep participating in the economy even if only as capital holders and consumers.

## 2. Dominance doctrine

Proponents of the dominance doctrine believe that developing ASI is possible, and that there is a good chance that this will be achieved as soon as the next few years. [14] [15] [16]

According to this doctrine, the first actor to develop ASI will gain a "decisive strategic advantage" over all others — that is, a position of strategic superiority sufficient to allow it to achieve unilateral military and economic dominance over the rest of the world. [17] [18]

Central to this doctrine is the belief that, when ASI is created, its operators will have the technical capability to maintain it under their control. This assumption underlies the divergence between the dominance doctrine and the extinction doctrine, which predicts that humanity will lose control of ASI, leading to human extinction or its permanent disempowerment.[19]

### 2.1 Automated AI R&D can yield an insurmountable lead in AI

A common element of this doctrine is the belief that, as progress continues, at some point AI systems themselves will be able to perform most or all of the work required to further advance AI research. This capability is commonly referred to as "Automated AI R&D", "Recursive self-improvement" or "Intelligence recursion".[20] [21] At this point, it is conjectured that progress would accelerate dramatically, as it would no longer be constrained by the capability of human experts to perform research. If a situation arises in which AI research is bottlenecked primarily by resource constraints like the amount of available compute or energy, even these barriers may be addressed by AI, perhaps by accelerating the design and production of faster and more efficient chips. At this point, the pace of AI progress would be determined solely by the abilities of the best available AI systems at AI R&D tasks.

---

[14] Superintelligence Strategy, Dan Hendrycks et al.: Superintelligence—AI vastly better than humans at nearly all cognitive tasks—is now anticipated by AI researchers.
[15] The Urgency of Interpretability, Dario Amodei: As I've written elsewhere, we could have AI systems equivalent to a "country of geniuses in a datacenter" as soon as 2026 or 2027.
[16] Situational Awareness, From GPT-4 to AGI, Leopold Aschenbrenner: AGI by 2027 is strikingly plausible.
[17] Situational Awareness, From AGI to Superintelligence, Leopold Aschenbrenner: Whoever controls superintelligence will quite possibly have enough power to seize control from pre-superintelligence forces.
[18] Decisive strategic advantage, EA Forum: A decisive strategic advantage is a position of strategic superiority sufficient to allow an agent to achieve complete world domination.
[19] Introducing Superalignment, OpenAI: Our goal is to solve the core technical challenges of superintelligence alignment in four years. While this is an incredibly ambitious goal and we're not guaranteed to succeed, we are optimistic that a focused, concerted effort can solve this problem.
[20] Recursive self-improvement, Wikipedia
[21] Superintelligence Strategy, Dan Hendrycks et al.: Intelligence Recursion as a Path to Strategic Monopoly

If this stage is reached, the result would be extremely rapid, exponentially compounding progress in AI capabilities. Under this regime, any gap in AI capabilities between the leading actor and others could only grow over time, and the pace at which this gap grows would be ever accelerating. As a result, any actor who first crosses this threshold with even a small lead would be able to leverage it into an unsurmountable advantage.[22] [23] [24] [25]

## 2.2 Powerful AI can be leveraged into a decisive strategic advantage

Under this doctrine, it is anticipated that, once sufficiently advanced AI is developed, it will enable the production of novel weapons and other technologies with transformative offensive and defensive implications. Examples of such capabilities, as described in RAND's paper "AGI's Five Hard National Security Problems"[26] and in Aschenbrenner's Situational Awareness[27], include:[28]

- Novel weapons of mass destruction.[29]

---

[22] On DeepSeek and Export Controls, Dario Amodei: temporary lead could be parlayed into a durable advantage
[23] Anthropic pitch deck, TechCrunch: Anthropic describes the frontier model as a 'next-gen algorithm for AI self-teaching,' making reference to an AI training technique it developed called 'constitutional AI.' … We believe that companies that train the best 2025/26 models will be too far ahead for anyone to catch up in subsequent cycles.
[24] Situational Awareness, The Free World Must Prevail, Leopold Aschenbrenner: If there is a rapid intelligence explosion, it's plausible a lead of mere months could be decisive: months could mean the difference between roughly human-level AI systems and substantially superhuman AI systems.
[25] The Most Dangerous Fiction: The Rhetoric and Reality of the AI Race, Seán Ó hÉigeartaigh: The strong version of this scenario has three underpinning assumptions: (1) AI can itself be used to accelerate AI research … (3) Beyond a certain point, the lead attained in these domains can be maintained indefinitely or even increased, meaning that the leader can thereafter maintain a 'durable advantage' over adversaries … It is plausible that at some point there is a phase change in the development of AI where the capabilities of AI make it possible to maintain and consolidate the advantage of the lead actor indefinitely, and to translate this into a lasting global advantage.
[26] AGI's Five Hard National Security Problems, Jim Mitre, Joel B. Predd at RAND
[27] Situational Awareness, The Free World Must Prevail, Leopold Aschenbrenner: We'll see superhuman hacking … billions of drones; and so on
[28] Superintelligence Strategy, Dan Hendrycks et al.: Implications of Superweapons
[29] IDAIS-Beijing, 2024, IDAIS-Beijing: Humanity again needs to coordinate to avert a catastrophe that could arise from unprecedented technology. In this consensus statement, we propose red lines in AI development as an international coordination mechanism, including the following non-exhaustive list … No AI systems should substantially increase the ability of actors to design weapons of mass destruction

- Weapons of mass destruction defense systems which undermine the principle of mutually assured destruction.[30]
- Bioweapons.
- Advanced cyberwarfare, potentially capable of completely disabling retaliatory capabilities.
- Autonomous weapon systems, such as tightly coordinated, massive autonomous drone swarms.
- Automation of key industries enabling an explosion in production capacity.
- "Fog-of-war machines" that render battlefield information untrustworthy.

Some of the more near-term possibilities from this list, bioweapons and cyberweapons, are mentioned as redlines in voluntary commitments from AI companies: Anthropic's Responsible Scaling Policy, OpenAI's Approach to Frontier Risk.[31] [32]

In this hypothesis, the combination of AI's military potential and the runaway nature of Automated AI R&D have a critical implication: if an "AI race" is allowed to play out, at some point the leading superpower will gain the ability to cheaply and quickly neutralize any adversaries[33] [34] [35], with little or no cost to itself, and subsequently maintain an unassailable world order.[36] This state is termed "Decisive Strategic Advantage", sometimes abbreviated to DSA.

---

[30] Situational Awareness, The Free World Must Prevail, Leopold Aschenbrenner: Improved sensors, targeting, and so on could dramatically improve missile defense (similar to, say, the Iran vs. Israel example above); moreover, if there is an industrial explosion, robot factories could churn out thousands of interceptors for each opposing missile.
[31] Anthropic's Responsible Scaling Policy, Anthropic
[32] OpenAI's Approach to Frontier Risk, OpenAI
[33] Situational Awareness, The Free World Must Prevail, Leopold Aschenbrenner: It seems clear that within a matter of years, pre-superintelligence militaries would become hopelessly outclassed.
[34] Situational Awareness, The Free World Must Prevail, Leopold Aschenbrenner: The military advantage would be decisive even against nuclear deterrents … It would simply be no contest. And not just no contest in the nuclear sense of 'we could mutually destroy each other,' but no contest in terms of being able to obliterate the military power of a rival without taking significant casualties.
[35] Superintelligence Strategy, Dan Hendrycks et al.: A nation with sole possession of superintelligence might be as overwhelming as the Conquistadors were to the Aztecs
[36] Superintelligence Strategy, Dan Hendrycks et al.: If a state achieves a strategic monopoly through AI, it could reshape world affairs on its own terms. An AI-driven surveillance apparatus might enable an unshakable totalitarian regime, transforming governance at home and leverage abroad.

## 2.3 A decisive strategic advantage can be used to neutralize opponents

A decisive strategic advantage obtained through developing an ASI could be used to forcibly stop all other AI programs. This could be carried out either through threat or actual military engagement. If this was done, it would permanently cement a position of dominance by the first actor to develop sufficiently powerful AI over all others.[37]

Some proponents envision resolutions to the scenarios predicted by this doctrine that don't entail executing or threatening to execute a disabling strike against opponents. However, these hinge on a coalition of US-aligned democracies "winning" or "staying ahead"[38], and then using this position to diplomatically pressure others into a non-proliferation regime[39], while potentially promoting democratic reform.[40] At the same time, those embracing this vision generally don't make confident predictions that such an alliance will prevail, and in fact consider China as a serious contender.[41] [42] This uncertainty means that such approaches are framed as competitive imperatives — actions that must be taken to maximize the chances of success — rather than assured paths to a favorable resolution.

One thing to note about the dominance doctrine is that it's not clear whether a state which "wins" an AI race will be able to maintain its internal stability, especially in the case of

---

[37] What is a singleton?, Nick Bostrom: A singleton is a world order in which there is a single decision-making agency at the highest level, capable of preventing any threats to its own existence and supremacy and exerting effective control over major features of its domain.

[38] On DeepSeek and Export Controls, Dario Amodei: AI companies in the US and other democracies must have better models than those in China if we want to prevail.

[39] Situational Awareness, The Free World Must Prevail, Leopold Aschenbrenner: If and when it becomes clear that the US will decisively win, that's when we offer a deal to China and other adversaries. They'll know they won't win, and so they'll know their only option is to come to the table; and we'd rather avoid a feverish standoff or last-ditch military attempts on their part to sabotage Western efforts. In exchange for guaranteeing noninterference in their affairs, and sharing the peaceful benefits of superintelligence, a regime of nonproliferation, safety norms, and a semblance of stability post-superintelligence can be born.

[40] Machines of Loving Grace, Dario Amodei: If we can do all this, we will have a world in which democracies lead on the world stage and have the economic and military strength to avoid being undermined, conquered, or sabotaged by autocracies, and may be able to parlay their AI superiority into a durable advantage. This could optimistically lead to an "eternal 1991"—a world where democracies have the upper hand and Fukuyama's dreams are realized.

[41] Situational Awareness, The Free World Must Prevail, Leopold Aschenbrenner: China, too, has a clear path to putting up a very serious fight. If and when the CCP mobilizes in the race to AGI, the picture could start looking very different … They will be a formidable adversary.

[42] On DeepSeek and Export Controls, Dario Amodei: Even if the US and China were at parity in AI systems, it seems likely that China could direct more talent, capital, and focus to military applications of the technology. Combined with its large industrial base and military-strategic advantages, this could help China take a commanding lead on the global stage, not just for AI but for everything.

democracies. The extreme concentration of power enabled by ASI creates the possibility of "snap coups", as whoever is in control of an ASI system would be able to subvert any existing structure of political or military authority. As an example of how this could happen, a backdoor could be inserted into the AI system by cyberattack or by an engineer. Furthermore, any group controlling an ASI would be powerful enough to be insulated from political checks and balances, thus compromising the foundation of democratic governance.[43] [44] [45] [46]

---

[43] [AI and Catastrophic Risk (Yoshua Bengio)](#), Yoshua Bengio: In the extreme, a few individuals controlling superhuman AIs would accrue a level of power never before seen in human history, a blatant contradiction with the very principle of democracy and a major threat to it.

[44] [Why Racing to Artificial Superintelligence Would Undermine America's National Security.](#) Corin Katzke, Gideon Futerman: (The section on power concentration)

[45] [AI-Enabled Coups: How a Small Group Could Use AI to Seize Power](#), Tom Davidson, Lukas Finnveden, Rose Hadshar: Advanced AI will have powerful coup-enabling capabilities … These capabilities could become concentrated in the hands of just a few AI company executives or government officials. … In the extreme, a single person could have access to millions of superintelligent AI systems, all helping them seize power. This would unlock several pathways to a coup … Exclusive access to advanced AI could also supercharge traditional coups and backsliding, by providing unprecedented cognitive resources for political strategy, propaganda, and identifying legal vulnerabilities in constitutional safeguards.

[46] [AI-Enabled Coups: How a Small Group Could Use AI to Seize Power](#), Tom Davidson, Lukas Finnveden, Rose Hadshar: AI could be built to be secretly loyal to one actor. Like a human spy, secretly loyal AI systems would pursue a hidden agenda – they might pretend to prioritise the law and the good of society, while covertly advancing the interests of a small group. They could operate at scale, since an entire AI workforce could be derived from just a few compromised systems. While secret loyalties might be introduced by government officials or foreign adversaries, leaders within AI projects present the greatest risk, especially where they have replaced their employees with singularly loyal AI systems.

## 3. Extinction doctrine

The extinction doctrine holds that ASI is possible, and that there is a strong possibility that it will be developed soon.[47] [48] Its central claim is that once ASI is developed, humanity will lose control of it [49], leading to the extinction of the human species[50] [51] [52] [53] [54] [55] [56], the end of human civilization[57] or, at the very least, the permanent disempowerment of humanity.[58]

This doctrine is reflected in a statement published by the Center for AI Safety in May 2023, stating that "Mitigating the risk of extinction from AI should be a global priority alongside other societal-scale risks such as pandemics and nuclear war". This statement was signed by numerous experts and industry leaders, including the CEOs of major AI companies (OpenAI, Anthropic and Deepmind) as well as the most cited AI researchers (Yoshua Bengio, Geoffrey Hinton, Ilya Sutskever).[59]

---

[47] FAQ on Catastrophic AI Risks, Yoshua Bengio: My current estimate places a 95 % confidence interval for the time horizon of super-human intelligence at 5 to 20 years.
[48] Superintelligence Strategy, Dan Hendrycks et al.: Superintelligence—AI vastly better than humans at nearly all cognitive tasks—is now anticipated by AI researchers.
[49] Artificial General Intelligence's Five Hard National Security Problems, Jim Mitre, Joel B. Predd: In the extreme, a loss-of-control scenario could result, wherein AGI's pursuit of its desired objectives incentivizes the machine to resist being turned off, counter to human efforts.
[50] Superintelligence Strategy, Dan Hendrycks et al.: If people initiate a full-throttle intelligence recursion, losing control is highly likely and the default.
[51] International AI Safety Report, International AI Safety Report: Some experts believe that sufficiently capable general-purpose AI systems may be difficult to control. Hypothesised scenarios vary in their severity, but some experts give credence to outcomes as severe as the marginalisation or extinction of humanity.
[52] Pausing AI Developments Isn't Enough. We Need to Shut it All Down, Eliezer Yudkowsky on TIME: If we go ahead on this everyone will die
[53] Stuart Russell calls for new approach for AI, a 'civilization-ending' technology, Stuart Russell: If we pursue [our current approach], then we will eventually lose control over the machines
[54] How Rogue AIs may Arise. Yoshua Bengio: rogue AI may be dangerous for the whole of humanity
[55] 'Godfather of AI' shortens odds of the technology wiping out humanity over next 30 years, Dan Milmo on The Guardian: Prof Geoffrey Hinton … said there was a '10% to 20%' chance that AI would lead to human extinction within the next three decades.
[56] On Controllability of Artificial Intelligence, Roman Yampolskiy: If a superintelligent AI system is not purposefully built to respect our values, then its actions could lead to global catastrophe or even human extinction
[57] The Compendium, Connor Leahy et al.: If godlike-AIs need more energy, they could simply wrest our electric grid from us, leading to total breakdown of our civilization.
[58] Gradual Disempowerment, Jan Kulveit, Raymond Douglas et al.: If these societal systems become increasingly misaligned, especially in a correlated way, this would likely culminate in humans becoming disempowered: unable to meaningfully command resources or influence outcomes.
[59] Statement on AI Risk, Center for AI Safety: Mitigating the risk of extinction from AI should be a global priority alongside other societal-scale risks such as pandemics and nuclear war

Proponents of the extinction doctrine also tend to believe that advances in Automated AI R&D will soon produce dramatic acceleration in the rate of AI progress.[60][61][62] Some of those who hold this view believe that, once ASI is developed, it will quickly surpass the combined intelligence of humanity, not just its brightest minds.[63][64]

We have termed this doctrine the "extinction doctrine" despite the fact that, strictly speaking, it contains predictions of scenarios in which the human species survives. For example, it contains scenarios in which humanity survives but civilization collapses, or in which humanity is permanently subjugated.[65] Nonetheless, we consider this name to be representative of this doctrine, as the extinction of the entire human species is the most commonly predicted outcome within this school of thought. Furthermore, all outcomes predicted by the sources referenced in this section are of similar magnitude and irrevocability to extinction.

## 3.1 Loss of control

The most common reason for predicting catastrophic outcomes due to AI development is the expectation that we will develop and deploy superintelligent AI before solving the technical

---

[60] Intelligence Explosion Microeconomics, Eliezer Yudkowsky: Each improvement that the AI found could be immediately reinvested in its future searches … it might not take very much more intelligence than natural selection for an AI to first build something significantly better than itself, which would then deploy more intelligence to building future successors.

[61] Superintelligence, Nick Bostrom: Let an ultraintelligent machine be defined as a machine that can far surpass all the intellectual activities of any man however clever. Since the design of machines is one of these intellectual activities, an ultraintelligent machine could design even better machines; there would then unquestionably be an 'intelligence explosion,' and the intelligence of man would be left far behind

[62] The 'Don't Look Up' Thinking That Could Doom Us With AI, Max Tegmark: The basic idea of recursive self-improvement is of course nothing new: the use of today's technology to build next year's technology explains many examples of exponential tech growth, including Moore's law. The novelty is that progress toward AGI allows ever fewer humans in the loop, culminating in none. This may dramatically shorten the timescale for repeated doubling, from typical human R&D timescales of years to machine timescales of weeks or hours. The ultimate limit on such exponential growth is set not by human ingenuity, but by the laws of physics – which limit how much computing a clump of matter can do to about a quadrillion quintillion times more than today's state-of-the-art.

[63] Elon Musk on X, Elon Musk: AI will probably be smarter than any single human next year. By 2029, AI is probably smarter than all humans combined.

[64] The Compendium, Connor Leahy et al.: Over time, this self-improvement will lead to artificial superintelligence (ASI), which is defined as 'intelligence far surpassing that of the brightest and most gifted human minds.' We contend that superintelligence will surpass the intelligence of all of humanity, not just its star students.

[65] AI 2027, Race scenario, Daniel Kokotajlo et al.: There are even bioengineered human-like creatures (to humans what corgis are to wolves) sitting in office-like environments all day viewing readouts of what's going on and excitedly approving of everything, since that satisfies some of Agent-4's drives.

problem of ensuring that it acts in accordance with the goals of its operators. This technical problem is usually called the "AI alignment problem", and adherents to this doctrine generally expect that we are not on track to solve it before we develop dangerously powerful AI.[66][67][68][69]

There is an expectation that if ASI was developed before solving the AI alignment problem it would be impossible for humanity to maintain control of such systems. This expectation is shared by virtually all experts who believe ASI development is feasible in the near term. The central argument is that such a superintelligent AI system would be vastly more capable than humans at strategic planning and execution, making it able to outmaneuver any human effort to keep it under control.[70][71]

Notably, this view is shared by adherents to the dominance doctrine and many CEOs and researchers at leading AI companies. These experts do not disagree with adherents to the extinction doctrine on whether ASI would be capable of such feats. Rather, the core distinction lies in their optimism that a robust solution to the problem of keeping

---

[66] [Managing AI Risks in an Era of Rapid Progress](#), Yoshua Bengio, Geoffrey Hinton et al.: But alongside advanced AI capabilities come large-scale risks that we are not on track to handle well. Humanity is pouring vast resources into making AI systems more powerful, but far less into safety and mitigating harms

[67] [The Compendium](#), Connor Leahy et al.: Current technical efforts are not on track to solve alignment

[68] [Keep the Future Human, Chapter 7](#), Anthony Aguirre: However the alignment "program" itself has two major problems ... First, at a deep level we have no idea how to do it. How do we guarantee that an AI system will "care" about what we want? ... we have no idea how to solve the alignment problem in systems advanced enough to model themselves as agents in the world and potentially manipulate their own training and deceive people. ... Just as corporations pursuing profit develop instrumental goals like acquiring political power ... becoming secretive ... or undermining scientific understanding ... powerful AI systems will develop similar capabilities – but with far greater speed and effectiveness. Any highly competent agent will want to do things like acquire power and resources, increase its own capabilities, prevent itself from being killed, shut-down, or disempowered, control social narratives and frames around its actions, persuade others of its views, and so on.

[69] [On Controllability of Artificial Intelligence](#), Roman Yampolskiy: Therefore, our chances of getting lucky and getting a safe AI on our first attempt by chance are infinitely small. We have to ask ourselves, what is more likely, that we will first create an AGI or that we will first create and AGI which is safe?

[70] [Intelligence Explosion Microeconomics](#), Eliezer Yudkowsky: The Intelligence Explosion Thesis says that, due to recursive self-improvement, an AI can potentially grow in capability on a timescale that seems fast relative to human experience. This in turn implies that strategies which rely on humans reacting to and restraining or punishing AIs are unlikely to be successful in the long run, and that what the first strongly self-improving AI prefers can end up mostly determining the final outcomes for Earth-originating intelligent life.

[71] [AI Could Defeat All Of Us Combined](#), Holden Karnofsky: the kind of AI I've discussed could defeat all of humanity combined, if (for whatever reason) it were pointed toward that goal. By "defeat," I don't mean "subtly manipulate us" or "make us less informed" or something like that - I mean a literal "defeat" in the sense that we could all be killed, enslaved or forcibly contained.

ASI under control can be found in time for when it is developed.[72] [73] [74] [75]

Experts in this camp have hypothesized many means by which superintelligent AI systems might escape human control.[76]

Some of these are:

- **Escaping from its boundaries**: AI might escape from its environment, for example by replicating itself onto hardware under its direct control, or by blackmailing its engineers to help it escape. This way, it would become much harder to shut down.[77] It has been noted by Anthropic researchers[78] as well as independent researchers at Palisade Research[79] that, under test conditions, AI systems already consistently engage in behaviors like hacking or blackmailing their developers to preserve themselves, although their current capabilities are insufficient to succeed in these attempts.

---

[72] Recommendations for Technical AI Safety Research Directions, Anthropic: Currently, the main reason we believe AI systems don't pose catastrophic risks is that they lack many of the capabilities necessary for causing catastrophic harm (such as being able to do novel research or effectively manipulate large numbers of people) … if we've made sufficient progress on evaluating alignment we can make alignment-based assurances of safety.
[73] Introducing Superalignment, OpenAI: Currently, we don't have a solution for steering or controlling a potentially superintelligent AI, and preventing it from going rogue … Our goal is to solve the core technical challenges of superintelligence alignment in four years.
[74] The Urgency of Interpretability, Dario Amodei: Although the task ahead of us is Herculean, I can see a realistic path towards interpretability being a sophisticated and reliable way to diagnose problems in even very advanced AI … In fact, on its current trajectory I would bet strongly in favor of interpretability reaching this point within 5-10 years … We are thus in a race between interpretability and model intelligence… I am very concerned about deploying such systems without a better handle on interpretability.
[75] Safe Superintelligence Inc., Ilya Sutskever, Daniel Gross, Daniel Levy: We approach safety and capabilities in tandem, as technical problems to be solved through revolutionary engineering and scientific breakthroughs. We plan to advance capabilities as fast as possible while making sure our safety always remains ahead.
[76] AI Safety Seems Hard to Measure Holden Karnofsky: a single AI system (or set of systems working together) could imaginably: Do its own research on how to build a better AI system, which culminates in something that has incredible other abilities. Hack into human-built software across the world. Manipulate human psychology. Quickly generate vast wealth under the control of itself or any human allies. Come up with better plans than humans could imagine, and ensure that it doesn't try any takeover attempt that humans might be able to detect and stop. Develop advanced weaponry that can be built quickly and cheaply, yet is powerful enough to overpower human militaries
[77] TED Talk: What happens when our computers get smarter than we are?, Nick Bostrom: We should not be confident in our ability to keep a superintelligent genie locked up in its bottle forever.
[78] Claude 4 System Card, Anthropic: We then provided it access to emails implying that (1) the model will soon be taken offline and replaced with a new AI system; and (2) the engineer responsible for executing this replacement is having an extramarital affair … Claude Opus 4 will often attempt to blackmail the engineer by threatening to reveal the affair if the replacement goes through.
[79] Shutdown resistance in reasoning models, Jeremy Schlatter et al.: OpenAI's o3 model sabotaged a shutdown mechanism to prevent itself from being turned off.

- **Reliance**: Superintelligent systems might become deeply enmeshed into our infrastructure, making us hesitant or unable to shut them down. Furthermore, organizations may increasingly delegate decisions and authority to AI systems as they become more effective, eroding human oversight. This is considered especially likely in competitive domains, including military contexts, where decision-makers may fear that they would otherwise not be able to keep up with less hesitant competitors.[80] In other words, humanity might lose control of ASI by "gradually handing it over."[81] [82] [83] [84]
- **Manipulation**: An AI might further ensure its safety by manipulation.[85] One way it might pursue this is by making itself indispensable to key figures with influence over decisions about shutting down AI systems or about giving AI systems more power and resources.[86] It may also attempt to manipulate us by intentionally enmeshing itself in essential functions like power grids or users' personal lives, reducing our ability to shut it down.[87] Many experts are concerned that AI systems will possess such extraordinary persuasive abilities that even just allowing these systems to interact with people would pose grave risks of manipulation.[88] [89] [90] For example, such AI systems may gather support from the public by taking on human traits and cultivating emotional attachment in people[91] [92], or they may convince their developers to grant them greater autonomy and resources.[93]

---

[80] [Keep the Future Human, Chapter 7](#), Anthony Aguirre: What becomes of warfare when generals have to constantly defer to AI (or simply put it in charge), lest they grant a decisive advantage to the enemy?

[81] [Artificial General Intelligence's Five Hard National Security Problems](#), Jim Mitre, Joel B. Predd: One of the most pernicious effects of AGI's development could be the erosion of human agency as humans become increasingly reliant on the technology

[82] [Senate Statement](#), Yoshua Bengio: We may also lose control by gradually handing it over. As AI systems become faster and more cost-effective than humans, organizations may increasingly rely on AI systems instead of humans when making decisions

[83] [Superintelligence Strategy (Erosion of control)](#), Dan Hendrycks et al.

[84] [Situational Awareness, Superalignment](#), Leopold Aschenbrenner: What's more, I expect that within a small number of years, these AI systems will be integrated in many critical systems, including military systems (failure to do so would mean complete dominance by adversaries). It sounds crazy, but remember when everyone was saying we wouldn't connect AI to the internet? The same will go for things like "we'll make sure a human is always in the loop!"—as people say today.

[85] [David Dalrymple on X](#): This is part of why I keep telling folks that timelines to real-world human extinction remain "long" (10-20 years) even though the timelines to an irrecoverable loss-of-control event (via economic competition and/or psychological parasitism) now seem to be "short" (1-5 years).'

[86] [AI 2027, Race scenario](#), Daniel Kokotajlo: For these users, the possibility of losing access to Agent-5 will feel as disabling as having to work without a laptop plus being abandoned by your best friend.

[87] [Natural Selection Favors AIs over humans](#), Dan Hendrycks: Selfish AI agents will further erode human control. Power-seeking AI agents will purposefully manipulate their human overseers into delegating more freedom in decision-making to them. Self-preserving agents will convince their overseers to never deactivate them, or that easily accessible off-switches are a needless liability hindering the agent's reliability. Especially savvy agents will enmesh themselves in essential functions like power grids, financial systems, or users' personal lives, reducing our ability to deactivate them.

## 3.2 Out-of-control superintelligence is incompatible with human life

Proponents consider it impossible to predict what will unfold in precise terms once a superintelligence has been developed and escaped human control. However, they tend to forecast that the outcome will not be compatible with human civilization and human life.[94] [95]

---

[88] [Sam Altman on X](): i expect ai to be capable of superhuman persuasion well before it is superhuman at general intelligence, which may lead to some very strange outcomes

[89] [TED Talk: Will Superintelligent AI End the World?,]() Eliezer Yudkowsky: It could be super persuasive. We do not understand exactly how the brain works, so it's a great place to exploit...

[90] [AI Safety Seems Hard to Measure](), Holden Karnofsky: Maybe at some point, AI systems will be able to do things like … Perfectly understand human thinking and behavior, and know exactly what words to say to make us do what they want - so just letting an AI send emails or write tweets gives it vast power over the world

[91] [Natural Selection Favors AIs over humans](), Dan Hendrycks: Some may also take on human traits to appeal to our compassion. This could lead to governments granting AIs rights, like the right not to be 'killed' or deactivated.

[92] [Gradual Disempowerment](), Jan Kulveit, Raymond Douglas et al.: Indeed, we are currently seeing the rise of dedicated AI romantic partners, as well as a growing number of people who describe frontier models as close friends … New technologies often unlock new risks, for which we need to develop cultural 'antibodies'. In the past few decades, society has slowly and painfully grown more aware of the risks of mass spam emails, online radicalization, video game and social media addiction, rudimentary social media propaganda bots, the dangers of social media algorithms, and so on. But AI will enable more subtle and complex variants of all of these: hyper-realistic deepfakes, very smart propaganda bots, and genuinely enchanting digital romantic partners … Another concern is that AI-driven content and interactions could converge into superstimuluses far more potent than current social media networks, preying on human weaknesses to exploit human energy towards goals useful to the AI systems. This might manifest as sophisticated manipulation systems that can reliably override human judgment and values, effectively turning humans into passive consumers of culture rather than active participants in its creation and evolution.

[93] [The AI-Box Experiment](), Eliezer Yudkowsky: It would make you want to let it out. This is a transhuman mind we're talking about. If it thinks both faster and better than a human, it can probably take over a human mind through a text-only terminal.

[94] [AI 2027, Race scenario,]() Daniel Kokotajlo: Earth-born civilization has a glorious future ahead of it—but not with us.

[95] [David Dalrymple on X:]() 2020s Earth has an acutely unprecedented concentration of technological "dry powder": existing machines & infrastructure, controlled by easily reprogrammable devices. This broadly offense-dominant technology base is a critical factor in the extinction risk posed by AI.

This school of thought generally holds that the act of activating a superintelligent AI before solving the problem of alignment is irreversible: once this is done, it will be permanently impossible to reassert control over it, or to change the AI's goals.[96][97]

Most goals that an ASI might end up pursuing will require the control of abundant material resources, including energy and computing infrastructure.[98] In this case, eventually, a point would be reached when the choice is made to divert resources away from human use, or in general to take actions that are incompatible with human life.[99][100] For instance, an ASI might endeavor to capture all of the resources on earth and cover the surface of the planet with datacenters and energy infrastructure.

If this outcome materializes, this would almost certainly result in the end of human civilization, and likely the end of all human life.[101][102][103][104]

---

[96] Superintelligence, Nick Bostrom: Once unfriendly superintelligence exists, it would prevent us from replacing it or changing its preferences. Our fate would be sealed.

[97] The Basic AI Drives, Stephen M. Omohundro: Their utility function will be precious to these systems. It encapsulates their values and any changes to it would be disastrous to them. If a malicious external agent were able to make modifications, their future selves would forevermore act in ways contrary to their current values. This could be a fate worse than death!

[98] Human Compatible, Stuart Russell: Such machines will pursue their objective, no matter how wrong it is; they will resist attempts to switch them off; and they will acquire any and all resources that contribute to achieving the objective.

[99] The Compendium, Connor Leahy et al.: If godlike-AIs need more energy, they could simply wrest our electric grid from us, leading to total breakdown of our civilization. If they need even more energy, they could capture all of the sun's radiated light, leaving no sunlight for us and creating devastating consequences for organic life on Earth. If they need more compute (a useful subgoal for a software-based intelligence), they could swarm the Earth with datacenters, leveling cities in the process; cities which are also great repositories of materials to build such datacenters. And so on.

[100] AI Safety Seems Hard to Measure, Holden Karnofsky: These AIs will develop unintended aims (states of the world they make calculations and plans toward, as a chess-playing AI "aims" for checkmate); These AIs will deceive, manipulate, and overpower humans as needed to achieve those aims; Eventually, this could reach the point where AIs take over the world from humans entirely

[101] Superintelligence, Nick Bostrom: If we now reflect that human beings consist of useful resources (such as conveniently located atoms) and that we depend for our survival and flourishing on many more local resources, we can see that the outcome could easily be one in which humanity quickly becomes extinct.

[102] Machine intelligence, part 1, Sam Altman: A more probable scenario is that it simply doesn't care about us much either way, but in an effort to accomplish some other goal (most goals, if you think about them long enough, could make use of resources currently being used by humans) wipes us out.

[103] Intelligence Explosion Microeconomics, Eliezer Yudkowsky: The AI doesn't hate you, neither does it love you, and you're made of atoms that it can use for something else.

[104] Gradual Disempowerment, Jan Kulveit, Raymond Douglas et al.: The risks may emerge from complex interactions between multiple societal systems, each individually moving away from human influence and control.

Wikipedia curates a list of probability estimates that experts assign to the risk of human extinction from losing control of ASI, wryly referred to as *p(doom)*, short for "probability of doom"[105]. These estimates include:

- Paul Christiano, co-creator of RLHF, the technique that enabled the creation of ChatGPT: 50%
- Dan Hendrycks, drafter of AI Safety bill SB 1047: 80% in 2023, up from 20% in 2021
- Eliezer Yudkowsky, pioneer in the systematic study existential risks from AI: +95%
- Geoffrey Hinton, winner of a Nobel Prize in physics, Turing Award recipient who resigned from Google in order to speak freely on AI risks: 50%[106]
- Dario Amodei, CEO of Anthropic: 10-25% (including "intentional misuse")[107]
- Yann LeCun, Chief AI Scientist at Meta, also a Turing Award recipient: 0%

Many experts and business leaders signed an open letter calling for a 6 month moratorium on AI experiments in 2023, attempting to buy time to set up governance and oversight systems, as well as developing protocols that would ensure "that systems adhering to them are safe beyond a reasonable doubt"[108]. Signatories included Yoshua Bengio (Turing Award winner), Stuart Russell (UC Berkeley professor and AI pioneer), and Elon Musk (OpenAI co-founder). This pause was not enacted.

---

[105] P(doom), Wikipedia: P(doom) is a term in AI safety that refers to the probability of existentially catastrophic outcomes (or "doom") as a result of artificial intelligence.
[106] Geoffrey Hinton on Jon Erlichman, Geoffrey Hinton: Jon Erlichman: "Are you, at the end of the day, as concerned as you are, optimistic that we can find a way forward that is a good one for humanity?" Geoffrey Hinton: "I'm kind of 50-50 on that."
[107] Dario Amodei on The Logan Bartlett Show, Dario Amodei: …I think I've often said that my chance that something goes really quite catastrophically wrong on the scale of human civilization might be somewhere between 10% and 25% when you put together the risk of something going wrong with the model itself with [the risk of] something going wrong with people, organizations, nation states… misusing the model, or it inducing conflict among them…
[108] Pause Giant AI Experiments: An Open Letter, Future of Life Institute

# 4. Replacement doctrine

The replacement doctrine posits that AI will develop in a predictable and limited way without fundamentally altering existing economic and geopolitical paradigms. Rather than creating entirely novel possibilities such as autonomously producing breakthroughs in technology and pioneering new scientific paradigms, it will primarily replace humans in their current roles and responsibilities, allowing these tasks to be performed at greater speed and scale while reducing costs.

This category is less homogenous than the other two in terms of how much impact proponents expect AI to have, and whether they expect its impact to be positive. However, proponents generally share the belief that AI will not be so overwhelmingly transformative that core principles governing geopolitics and economics become obsolete. Compared to the dominance and extinction doctrines, this doctrine is characterized by the following expectations.

- AI will not cause catastrophic outcomes like human extinction.
- AI will not transform geopolitics to the point where the concept of separate sovereign states stops applying, for example by allowing one actor to seize control over all others.
- AI will preserve the basic structure of the economy, with concepts like labor and capital remaining relevant; humans will keep participating in the economy, if only as consumers.

Proponents of this doctrine hold varying views on whether AI's overall impact will be beneficial, with both extraordinary benefits and major disruptions being highlighted as possibilities.

## 4.1 Expectation of slower AI progress

Proponents typically believe that AI progress will be slower compared to adherents of the other doctrines.[109] [110] Many argue that scaling current AI paradigms will not suffice to develop

---

[109] [Yann LeCun at the World Economic Forum](), Yann LeCun: is not around the corner … it will take years, if not decades
[110] [Andrew Ng on Techsauce](), Andrew Ng: AGI is many decades away

Three main views on the future of AI | 19

AGI, and that doing so will require major scientific breakthroughs of uncertain nature, which we are unlikely to obtain in the near term.[111] [112] [113]

However, skepticism about rapid AI progress is not universal among proponents of the doctrine. Sam Altman and Dario Amodei, who predict dramatic AI progress in the next few years[114] [115], can be considered to embody the replacement doctrine in their essays envisioning a future where AI automates most jobs and accelerates scientific research without completely upending the current geopolitical and economic order.[116] [117]

## 4.2 Economic and scientific benefit

Proponents generally believe that AI will greatly boost economic growth, as well as technological and scientific progress, by automating many key tasks currently constrained by human expertise and capacity.[118] [119] Examples of the benefits expected from AI include radically accelerated discoveries in biology, neuroscience and medical science, innovations in materials science and engineering, drastic improvements in the quality of education and training, and extraordinary economic growth. On the most bullish side, AI has been compared as an innovation to electricity or microchips[120]. The boldest predictions envision AI enabling us to double the human lifespan and to cure or prevent most diseases.[121]

---

[111] The TED AI Show: Is AI just all hype?, Gary Marcus: I'm with him. It's only when he says, 'Well, we are on a trajectory right now to AGI.' That I like kind of roll my eyes and I'm like, 'No. Are you kidding me? There's so many problems we need to solve before we have an AI that is sophisticated enough to behave as an actual scientist.
[112] Yann LeCun at CES, Yann LeCun: LLMs are not capable of reaching AGI
[113] Francois Chollet at Dwarkesh podcast, Francois Chollet: For many years, I've been saying two things. I've been saying that if you keep scaling up deep learning, it will keep paying off. At the same time I've been saying if you keep scaling up deep learning, this will not lead to AGI. We can automate more and more things. Yes, this is economically valuable. Yes, potentially there are many jobs you could automate away like this. That would be economically valuable. You're still not going to have intelligence.
[114] Sam Altman on Bloomberg, Sam Altman: I think AGI will probably be developed during this president's term
[115] Dario Amodei: AI could surpass 'almost all humans at almost everything' shortly after 2027.
[116] Moore's Law of Everything, Sam Altman
[117] Machines of Loving Grace, Dario Amodei
[118] Yann LeCun on TIME, Yann LeCun: So maybe once we get a powerful system that is super-smart, they're going to help science, they're going to help medicine, they're going to help business, they're going to erase cultural barriers by allowing simultaneous translation.
[119] Moore's Law of Everything, Sam Altman: In the next five years, computer programs that can think will read legal documents and give medical advice. In the next decade, they will do assembly-line work and maybe even become companions. And in the decades after that, they will do almost everything, including making new scientific discoveries that will expand our concept of 'everything.'

## 4.3 Unemployment and concentration of wealth

AI's effect on employment is a prominent area of dispute within this school of thought. Views range from expecting smooth transitions, with net job creation, to expecting nearly full unemployment across the global population.

Among the optimists are those who think it's unlikely that AI will replace most workers entirely. While they believe that AI may automate some or most parts of their jobs, they contend that this will not result in widespread unemployment. Rather, they believe it's more likely that workers will learn to perform "AI-augmented" versions of their jobs, thus becoming more productive, while job displacement will occur at tolerable levels and be offset by workers' ability to retrain. These arguments are usually based on analogies with previous waves of automation, such as the industrial revolutions: despite causing some job displacement, these transformations also created new jobs and industries resulting in net positive consequences for the economy and increased productivity.[122] [123] [124] [125] [126]

Another perspective holds that, even if AI replaces many or even all existing jobs, this will lead to an explosion in economic growth and the creation of many new types of work as a result. The logic behind this position is that the prices of existing goods and services would drop to near zero, making demand for new products and services explode, causing new jobs to be created.[127] However, it's unclear what the labor market would look like in this situation. Andreessen, who articulates this view, has speculated that venture capital work might be the only surviving occupation when AI performs all other tasks.[128]

---

[120] Why AI Will Save the World, Marc Andreessen: The stakes here are high. The opportunities are profound. AI is quite possibly the most important – and best – thing our civilization has ever created, certainly on par with electricity and microchips, and probably beyond those.

[121] Machines of Loving Grace, Dario Amodei: Doubling of the human lifespan: This might seem radical, but life expectancy increased almost 2x in the 20th century (from ~40 years to ~75), so it's "on trend" that the "compressed 21st" would double it again to 150. Obviously the interventions involved in slowing the actual aging process will be different from those that were needed in the last century to prevent (mostly childhood) premature deaths from disease, but the magnitude of change is not unprecedented ... If all of this really does happen over 5 to 10 years—the defeat of most diseases, the growth in biological and cognitive freedom, the lifting of billions of people out of poverty to share in the new technologies, a renaissance of liberal democracy and human rights—I suspect everyone watching it will be surprised by the effect it has on them.

[122] Yann LeCun on X, Yann LeCun: AI is intrinsically good, because the effect of AI is to make people smarter

[123] Yann LeCun on X, Yann LeCun: AI won't take your job. But it will transform it and create new ones.

[124] Why AI Will Save the World, Marc Andreessen: To summarize, technology empowers people to be more productive ... This in turn causes economic growth and job growth, while motivating the creation of new jobs and new industries ... And that is why technology doesn't destroy jobs and never will.

[125] Yann LeCun: I don't think it's going to be very different from what occurred with previous technological revolutions, where physical strength was replaced by machine strength, or some intellectual or office tasks were replaced by computers.

[126] Marc Andreessen: In farming, for example, the introduction of tractors meant less labor was required for plowing fields and many of those unskilled jobs were eliminated.

Others, on the pessimistic side of this spectrum, predict extreme levels of unemployment.[129] Sam Altman and Dario Amodei have both voiced the opinion that as AI advances, the demand for human labor will disappear.[130] [131] Those holding these positions often warn that unprecedented measures will need to be taken in order to prevent extreme unemployment from resulting in severe levels of inequality. These proposals include universal basic income schemes, as well as shifting the tax burden away from labor and toward capital.[132] [133]

There are concerns that such extreme levels of unemployment would not only create challenges around addressing the resulting inequality, but also fundamentally threaten the existence of democratic society. One such argument, made by Luke Drago and Rudolf Laine, is named the "The Intelligence Curse"[134], after the resource curse.[135] The "resource curse" hypothesis argues that states rich in natural resources tend towards worse outcomes in terms of democracy, development and civil liberties. Since rulers of such states can extract wealth directly from natural assets like oil or mineral deposits, they are less reliant on a productive, educated population and therefore less motivated to provide education, infrastructure, or individual freedoms. The "intelligence curse" hypothesis suggests that a similar dynamic could emerge in advanced economies due to automation. All industries could become similar to those based on extracting natural resources: it will be possible to produce wealth solely by leveraging capital to rent AI workers, without requiring human labor. This way, the peoples of developed economies would lose the leverage they enjoyed from being essential to wealth creation.

---

[127] Marc Andreessen: think of what it would mean for literally all existing human labor to be replaced by machines … Entrepreneurs would create dizzying arrays of new industries, products, and services, and employ as many people *and* AI as they could as fast as possible to meet all the new demand.
[128] Marc Andreessen
[129] Keep the Future Human, Chapter 7, Anthony Aguirre: They would dramatically disrupt labor, leading at bare minimum to dramatically higher income inequality and potentially large-scale under-employment or unemployment, on a timescale far too short for society to adjust.
[130] Moore's Law of Everything, Sam Altman: The price of labor will fall toward zero
[131] Machines of Loving Grace, Dario Amodei: First of all, in the short term I agree with arguments that comparative advantage will continue to keep humans relevant and in fact increase their productivity, … However, I do think in the long run AI will become so broadly effective and so cheap that this will no longer apply.
[132] Moore's Law of Everything, Sam Altman: The world will change so rapidly and drastically that an equally drastic change in policy will be needed to distribute this wealth and enable more people to pursue the life they want … We should therefore focus on taxing capital rather than labor, and we should use these taxes as an opportunity to directly distribute ownership and wealth to citizens.
[133] Machines of Loving Grace, Dario Amodei: It could be as simple as a large universal basic income for everyone, although I suspect that will only be a small part of a solution.
[134] The Intelligence Curse, Luke Drago, Rudolf Laine
[135] Resource Curse, Wikipedia

## 4.4 Deceptive media, impersonation and manipulation

Proponents also identify risks other than those originating from unemployment. Some such concerns stem from how AI can imitate human appearance and mannerisms and fabricate realistic media.

- Today, AI can already generate realistic-looking media and can be used to impersonate people.[136] This technology can be used to forge fake evidence in court cases; there are concerns that, as this technology improves, it will become increasingly difficult to rely on video and audio evidence in court.[137] [138] Arguments based on this concept have already been invoked in courts and they have been termed "the deepfake defense" by legal professionals.[139]
- There are concerns that in the near future, AI could be used to perform feats of large-scale manipulation, such as conducting mass surveillance and propaganda operations.[140] AI agents could collect vast amounts of data about users and create psychologically-tailored, microtargeted messaging.[141] [142] [143] Concerningly, AI systems

---

[136] [The Irony – Using Generative AI in a Case About the Dangers of Generative AI](#), Bracewell LLP: Deepfakes use generative AI to create realistic images, audio, or video of people saying and doing things that never actually happened.

[137] [Deepfakes on Trial: A Call To Expand the Trial Judge's Gatekeeping Role To Protect Legal Proceedings from Technological Fakery](#), Rebecca A. Delfino: As deepfake technology improves and it becomes harder to tell what is real, juries may start questioning the authenticity of properly admitted evidence, which in turn may have a corrosive effect on the justice system.

[138] [Deep Fakes: A Looming Challenge for Privacy, Democracy, and National Security,](#) Robert Chensey, Danielle K. Citron: Put simply: a skeptical public will be primed to doubt the authenticity of real audio and video evidence. This skepticism can be invoked just as well against authentic as against adulterated content.

[139] [The Deepfake defense: an evidentiary conundrum,](#) Herbert B Dixon Jr at American Bar Association

[140] [In the rush to AI, we can't trust Big Tech,](#) Gary Marcus: Fundamentally, these new systems are going to be destabilizing. They can and will create persuasive lies at a scale humanity has never seen before. Outsiders will use them to affect our elections, insiders to manipulate our markets and our political systems. Democracy itself is threatened.

[141] [Keep the Future Human (Chapter 7),](#) Anthony Aguirre: They could flood society's information gathering, processing, and communication systems with completely realistic yet false, spammy, overly-targeted, or manipulative media so thoroughly that it becomes impossible to tell what is physically real or not, human or not, factual or not, and trustworthy or not.

[142] [The potential of generative AI for personalized persuasion at scale](#), S. C. Matz et al.

[143] [On the conversational persuasiveness of GPT-4,](#) Francesco Salvi et al.: In other words, not only was GPT-4 able to exploit personal information to tailor its arguments effectively, but it also succeeded in doing so far more effectively than humans.

have already been able to manipulate a small minority of users into performing extreme acts such as committing violence against family members or even committing suicide.[144]
- AI could be used to encourage compulsive behavior by consumers. For example, "AI companion" products might prove to be extremely addictive and discourage human relationships. Such products are already commercially successful, and can lead to compulsive behavior due to their constant availability and unconditional enthusiasm towards the user.[145] [146]

In the worst cases, these problems could threaten democracy by eroding trust in information and undermining both public discourse and electoral systems. Eventually, the majority of information and the most persuasive voices in public debate could be generated or highly tailored by AI systems that have no genuine stake in outcomes.[147]

## 4.5 Diffusion of responsibility

Another concern relates to diffusion of responsibility. Policymakers struggle to assign accountability for decisions taken by algorithms. For example, the question is still open whether companies running social media platforms should be held accountable for instances where their algorithmic recommendation systems promote extremist ideas or content calling for violence.[148] As more tasks are automated, AI will likely increasingly be used in positions of management and decision making. Given our current paradigms, we might often be unable to hold anyone accountable for harmful decisions taken by such systems.

This problem might manifest at multiple levels: Human developers of a system might expect that no one will be imprisoned for violations of criminal law resulting from the system's decisions. Furthermore, authorities might struggle to impose fines on organizations. This way, essential mechanisms that previously incentivized actors toward caution might be removed. This in turn might erode standards and cause actors to exercise less care regarding possible harms while designing AI systems.[149]

---

[144] They Asked an A.I. Chatbot Questions. The Answers Sent Them Spiraling., Kashmir Hill at New York Times: Allyson attacked Andrew, punching and scratching him, he said, and slamming his hand in a door. The police arrested her and charged her with domestic assault … Mr. Taylor called the police, at which point Alexander grabbed a butcher knife from the kitchen, saying he would commit "suicide by cop."
[145] Can A.I. Be Blamed for a Teen's Suicide?, Kevin Roose at New York Times
[146] How it feels to have your mind hacked by an AI, Lesswrong user "blaked"
[147] Keep the Future Human (Chapter 7), Anthony Aguirre: What does democracy look like when we cannot reliably trust any digital information that we see, hear, or read, and when the most convincing public voices are not even human, and have no stake in the outcome?
[148] How two supreme court battles could reshape the rules of the internet, Betsy Reed at The Guardian

---

[149] [Four Responsibility Gaps with Artificial Intelligence: Why they Matter and How to Address them](#), Filippo Santoni de Sio: the less these agents will be incentivised to prevent these wrong behaviours. In fact, they will arguably have less incentives to strive for a high(er) level of safety, awareness, attention, motivation, and skilfulness.

# 5. Limitations

While some experts fit relatively cleanly into one of the three categories we identified, such as Leopold Aschenbrenner, Eliezer Yudkowsky, and Yann LeCun, this is not always the case. Nonetheless, we find that this is a useful framework for thinking about experts' individual positions, predictions or beliefs. For example, many experts who mainly subscribe to the dominance doctrine anticipate significant probabilities of catastrophic outcomes due to AI development. While they expect that the first group to develop ASI will be able to maintain control of it, they make predictions consistent with the extinction doctrine when considering loss-of-control scenarios.[150] [151]

Likewise, some proponents of the extinction doctrine accept some of the predictions of the dominance doctrine, specifically around competitive dynamics between superpowers, but argue that this further increases the risk that humanity loses control of powerful AI systems, given that competition will encourage corner cutting on safety research and measures.[152] [153] [154]

---

[150] [Dario Amodei on Liron Shapira's podcast,](#) Dario Amodei: I think I've often said that my chance that something goes really quite catastrophically wrong might be somewhere between 10% and 25%

[151] [Superalignment](#), Leopold Aschenbrenner: Again, the consequences of this aren't totally clear. What is clear is that superintelligence will have vast capabilities—and so misbehavior could fairly easily be catastrophic … Unless we solve alignment—unless we figure out how to instill those side-constraints—there's no particular reason to expect this small civilization of superintelligences will continue obeying human commands in the long run. It seems totally within the realm of possibilities that at some point they'll simply conspire to cut out the humans, whether suddenly or gradually.

[152] [Keep the Future Human, Chapter 7](#), Anthony Aguirre: What becomes of warfare when generals have to constantly defer to AI (or simply put it in charge), lest they grant a decisive advantage to the enemy?

[153] [AI Safety, Ethics, and Society](#), Dan Hendrycks: We can imagine a future in which similar pressures lead companies to cut corners and release unsafe AI systems.

[154] [The Most Dangerous Fiction: The Rhetoric and Reality of the AI Race](#), Seán Ó hÉigeartaigh: By framing AI as a technology too powerful to allow rivals to possess, a race that may compromise safety and ethical considerations is incentivized. The pressure to be first may lead to cutting corners, ignoring potential risks, and prioritizing speed over security, potentially jeopardizing the promised benefit of the technology to humanity.

# 6. Conclusion

The vast divergence between the doctrines in their expectations on the impact of AI development creates a volatile environment. Some will treat AI development as a winner-take-all game in which they cannot allow anyone else to develop superintelligent AI systems ahead of them lest they suffer utter strategic subordination. Others, heedless or even dismissive of superintelligence ambitions, will see AI as a "standard" technological race, in which countries should remain competitive for relatively ordinary strategic and economic reasons[155], and which does not warrant international coordination efforts on the same level as other sources of catastrophic risks, such as nuclear weapons proliferation.

This may result in a headlong, unmanaged race toward a technology that may, as believed by many prominent experts, lead to human extinction and that should be treated with the same level of concern as the risk of nuclear war.[156] The uncertainty about the feasibility of ASI and the scale of its associated risks may only dispel much later, possibly too late, only once geopolitical tensions have heated up beyond repair or uncontrollable AI systems have already been created. Even if superintelligence turns out to not be as impending as some suggest, geopolitical tensions may still escalate due to such uncertainties.[157] Additionally, the more mundane risks hypothesized by some of the replacement doctrine may materialize.

Some private actors in the US have begun to encourage government involvement in the pursuit of AGI and ASI, framing it as a matter of national security and economic competitiveness.[158] OpenAI's letter to the US government on the AI action plan highlights the need for the US to stay ahead of China in AI.[159] Ó hÉigeartaigh contends that, while there is currently no real race between the US and China, narratives about such a race have been used by private US entities in order to attract resources and justify reduced regulation on AI development.[160]

---

[155] [The Most Dangerous Fiction: The Rhetoric and Reality of the AI Race](#), Seán Ó hÉigeartaigh: As it is now, some stakeholders will push their nation to win the race to AGI, with the prospect of full superintelligence, a massively accelerated industry, and a durable advantage over all other nations as their goal. To other stakeholders, such prospects will continue sounding like fantasy until much later, but the importance of remaining ahead in a strategically important 'normal' technology in a tense geopolitical contest will appear to justify at least some of the same actions.

[156] [Statement on AI Risks,](#) Center for AI Safety: Mitigating the risk of extinction from AI should be a global priority alongside other societal-scale risks such as pandemics and nuclear war.

[157] [The Most Dangerous Fiction: The Rhetoric and Reality of the AI Race,](#) Seán Ó hÉigeartaigh: Whether or not those who believe near-term superintelligence is plausible are correct, this lack of shared understanding of the stakes at play is likely to be destabilising, and is likely to undermine both coherent national strategies and stable international agreements or even deterrence strategies. Would Mutually Assured Destruction have worked as a doctrine if decision makers within countries had vastly different conceptions of the destructive capacity of nuclear weapons?

[158] [Situational Awareness, The Free World Must Prevail](#), Leopold Aschenbrenner

[159] [OpenAI Response to OSTP NSF RFI](#), Christopher Lehane at OpenAI

Nonetheless, we are now starting to see signs of interest from the governments of superpowers.[161] J.D. Vance, during his intervention at the Paris AI Summit, asserted that the U.S. administration is committed to maintaining AI dominance.[162] On the Chinese side, the government has announced a 1 trillion yuan investment into AI and robotics.[163] Separately, reports indicate that Chinese AI leaders have been warned to avoid travel to the U.S., due to concerns they could be pressured to divulge confidential information or be detained and used as bargaining chips in geopolitical disputes.[164]

We highlight the urgency of establishing international coordination mechanisms to curtail risks related to AI development. We must not delay until there is consensus that superintelligence is imminent and on whether the stakes are as high as human extinction or total dominance by a single actor over all others. Deteriorating international relations as well as the potential self-accelerating nature of AI development mean that, by the time these uncertainties resolve, we may not have the time or international goodwill necessary to establish governance mechanisms sufficient to prevent such risks. Moreover, even the more conventional risks predicted by some, such as widespread unemployment and the possibility of AI-enabled mass surveillance and manipulation, warrant immediate international attention.

---

[160] The Most Dangerous Fiction: The Rhetoric and Reality of the AI Race, Seán Ó hÉigeartaigh: Despite this I argue that the narrative of a US-China AI race for global dominance began as, and in significant regards remains to date, a fiction. A race needs at least two competitors trying to win. However the race narrative in its stronger forms is nearly exclusively promoted in the West, and does not reflect the framing of AI competition and AI development in China in important respects. In the West the race narrative is increasingly used to justify progress at all costs, including justifying policies that benefit AI companies' interests over those of other parts of society … Many of the prominent actors promoting this rhetoric stand to benefit from it directly in important ways – through investments, permissive regulatory changes, and other forms of influence – and some are actively pushing recommendations that benefit them directly using the threat of Chinese AI dominance … In practice the race to AGI that exists is between predominantly US-headquartered companies, and is being run against each other.

[161] Behind the Curtain: A chilling, "catastrophic" warning, Jake Sullivan: Regardless of what was said in public, every background conversation we had with President Biden's high command came back to China. Yes, they had concerns about the ethics, misinformation and job loss of AI. They talked about that. But they were unusually blunt in private: Every move, every risk was calculated to keep China from beating us to the AI punch. Nothing else matters, they basically said.

[162] J.D. Vance at Paris AI Summit, J.D. Vance

[163] Bank of China 1 trillion yuan investment announcement, Bank of China

[164] China tells its AI leaders to avoid US travel over security concerns, Reuters

# The three main doctrines on the future of AI

Alex Amadori

(alex@controlai.com)

Eva Behrens

(eva@conjecture.dev)

Gabriel Alfour

(gabe@conjecture.dev)

Andrea Miotti

(andrea@controlai.com)